\documentclass[final,journal]{IEEEtran} 

\usepackage{amssymb,graphicx,color,cite}
\newtheorem{defn}{Definition}
\newcommand{\binom}[2]{{{#1}\choose{#2}}}

\title{Staircase Codes: FEC for 100 Gb/s OTN\thanks{B. P. Smith and F.
R. Kschischang are with the Electrical and Computer Engineering
Department, University of Toronto, 10 King's College Road, Toronto,
Ontario M5S 3G4, Canada  (e-mail: \{ben, frank\}@comm.utoronto.ca).
\newline \indent A. Farhood is with Cortina Systems Inc., 535 Legget
Drive, Suite 1000, Kanata, Ontario K2K 3B8, Canada. \newline \indent
A. Hunt and J. Lodge are with the Communications Research Centre
Canada, 3701 Carling Ave., Ottawa, Ontario K2H 8S2, Canada.}}

\author{Benjamin P. Smith, Arash Farhood, Andrew Hunt,\\
Frank R.  Kschischang \IEEEmembership{Fellow, IEEE} and John Lodge
\IEEEmembership{Fellow, IEEE}} 

\IEEEpubid{0000--0000/00\$00.00~\copyright~2011 IEEE}

\begin{document}

\maketitle

\begin{abstract}\boldmath
Staircase codes, a new class of forward-error-correction (FEC) codes
suitable for high-speed optical communications, are introduced.  An
ITU-T G.709-compatible staircase code with rate $R=239/255$ is
proposed, and FPGA-based simulation results are presented, exhibiting
a net coding gain (NCG) of $9.41$~dB at an output error rate of
$10^{-15}$, an improvement of $0.42$~dB relative to the best code from
the ITU-T G.975.1 recommendation.  An error floor analysis technique
is presented, and the proposed code is shown to have an error floor at
$4.0 \times 10^{-21}$.
\end{abstract}


\section{Introduction}

\IEEEPARstart{A}{dvances} in physics---the invention of the laser,
low-loss optical fiber, and the optical amplifier---have driven the
exponential growth in worldwide data communications.  However, as
these technologies mature, system designers have increasingly focused
on techniques from communication theory, including forward error
correction, to simultaneously increase transmission capacity and
decrease transmission costs.

One of the first proposals for FEC in an optical system appeared
in~\cite{Gro1988}, which demonstrated a shortened $(224,216)$ Hamming
code implementation at 565 Mbit/s. Since then, ITU-T Recommendations
G.975 and G.975.1 have standardized more powerful codes for optical
transport networks (OTNs). More recently, low-density parity-check
(LDPC) codes~\cite{gal-book,RU2008a}---which provide the potential for
capacity-approaching performance---have been investigated, as aptly
summarized in~\cite{DAM2009a,M2006a}.  While implementations exists at
10 Gb/s (for 10GBase-T ethernet networks), the blocklengths of such
implementations ($\sim 500$--$2000$) are too short to provide
performance close to capacity; the $(2048,1723)$ RS-LDPC code is
approximately 3 dB from the Shannon Limit at
$10^{-15}$~\cite{zh2010a}, see also~\cite{DCK2008a}.  Another
significant roadblock is that fiber-optic communication systems are
typically required to provide bit-error-rates below $10^{-15}$. It is
well-known that capacity-approaching LDPC codes exhibit error
floors~\cite{rich2003}, and to achieve the targeted error rate would
likely require concatenation with an outer code (e.g., as
in~\cite{Miz2009a}).  In this work, we focus on product-like codes (by
product-like codes, we mean any generalized LDPC code with algebraic
component codes), since they possess properties that make them
particularly suited to providing error-correction in fiber-optic
communication systems. In particular, for 100 Gb/s implementations, we
argue that syndrome-based decoding of product-like codes is
significantly more efficient than message-passing decoding of LDPC
codes.  

This paper presents a new class of high-rate binary error-correcting
codes---staircase codes---whose construction combines ideas from
convolutional and block coding. Indeed, staircase codes can be
interpreted as having a `continuous' product-like construction.  In
the context of wireless communications, related code constructions
include braided block codes~\cite{fel2009}, braided convolutional
codes~\cite{zhan2010}, diamond codes~\cite{bagg} and cross parity
check convolutional codes~\cite{fuja}, each of which is related to the
recurrent codes of Wyner-Ash~\cite{wyn63}. However, these proposals
considered soft decoding of the component codes, which is unsuitable
for high-speed fiber-optic communications. Herein, we describe a
syndrome-based decoder for staircase codes, that provides excellent
performance with an efficient decoder implementation.

In Section~\ref{sec:exist}, we review the specifications and
performance of FEC codes defined in ITU-T Recommendations G.975 and
G.975.1.  In Section~\ref{sec:vs}, we describe the syndrome-based
decoder for product-like codes, and argue that it results in a decoder
data-flow that is more than two orders of magnitude smaller than the
message-passing decoder of an LDPC code.  Staircase codes are
presented in Section~\ref{sec:sc}, and a G.709-compatible staircase
code is proposed. In Section~\ref{sec:efloor}, we present an
analytical method for determining the error floor of iteratively
decoded staircase codes, and show that the proposed staircase code has
an error floor at $4.0\times 10^{-21}$. Finally, in
Section~\ref{sec:sim}, we present FPGA-based simulation results,
illustrating that the proposed code provides a $9.41$~dB NCG at an
output error rate of $10^{-15}$, an improvement of $0.42$~dB relative
to the best code from the ITU-T G.975.1 recommendation, and only
$0.56$ dB from the Shannon Limit. 

\IEEEpubidadjcol

\section{Existing Proposals} \label{sec:exist}
\subsection{ITU-T Recommendation G.975}

The first error-correction code standardized for optical
communications was the $(255,239)$ Reed-Solomon code, with symbols in
$\mathbb{F}_{2^8}$, capable of correcting up to $8$ symbol errors in
any codeword. For an output-error-rate of $10^{-15}$, the NCG of the
RS code is $6.2$ dB, which is $3.77$ dB from capacity.

In order to provide improved burst-error-correction, 16 codewords are
block-interleaved, providing correction for bursts of as many as 1024
transmitted bits.  A framing row consists of $16\cdot255\cdot8$ bits,
30592 of which are information bits, and the remaining 2048 bits of
which are parity.  The resulting framing structure---a frame consists
of four rows---is standardized in ITU-T recommendation G.709, and
remains the required framing structure for OTNs; as a direct result,
the coding rate of any candidate code must be $R=239/255$. 
 
\subsection{ITU-T Recommendation G.975.1}

As per-channel data rates increased to $10$ Gb/s, and the capabilities
of high-speed electronics improved, the $(255,239)$ RS code was
replaced with stronger error-correcting codes. In ITU-T recommendation
G.975.1, several `next-generation' coding schemes were proposed; among
the many proposals, the common mechanism for increased coding gain was
the use of concatenated coding schemes with iterative hard-decision
decoding. We now describe four of the best proposals, which will
motivate our approach in Section~\ref{sec:sc}.

In Appendix I.3 of G.975.1, a serially concatenated coding scheme is
described, with outer $(3860,3824)$ binary BCH code and inner
$(2040,1930)$ binary BCH code, which are obtained by shortening their
respective mother codes.  First, $30592=8\cdot3824$ information bits
are divided into 8 units, each of which is encoded by the outer code;
we will refer to the resulting unit of $30880$ bits as a `block'.
Prior to encoding by the inner code, the contents of consecutive
blocks are interleaved (in a `continuous' fashion, similar to
convolutional interleavers~\cite{for1971}). Specifically, each inner
codeword in a given block involves `information' bits from each of the
eight preceding `outer' blocks. Note that the interleaving step
increases the effective block-length of the overall code, but it
necessitates a sliding-window style decoding algorithm, due to the
continuous nature of the interleaver.  Furthermore, unlike a product
code, the parity bits of the inner code are protected by a single
component codeword, which reduces their level of protection. For an
output-error-rate of $10^{-15}$, the NCG of the I.3 code is $8.99$ dB,
which is $0.98$ dB from capacity.

In Appendix I.4 of G.975.1, a serially concatenated scheme with
(shortened versions of) an outer $(1023,1007)$ RS code and (shortened
versions of) an inner $(2047,1952)$ binary BCH code is proposed.
After encoding $122368$ bits with the outer code, the coded bits are
block interleaved and encoded by the inner BCH code, resulting in a
block length of $130560$ bits, i.e., exactly one G.709 frame. As in
the previous case, the parity bits of the inner code are
singly-protected. For an output-error-rate of $10^{-15}$, the NCG of
the I.4 code is $8.67$ dB, which is $1.3$ dB from capacity.

In Appendix I.5 of G.975.1, a serially concatenated scheme with an
outer $(1901,1855)$ RS code and an inner $(512,502)\times(510,500) $
extended-Hamming product code is described.  Iterative decoding is
applied to the inner product code, after which the outer code is
decoded; the purpose of the outer code is to eliminate the error floor
of the inner code, since the inner code has small stall patterns (see
Section~\ref{sec:efloor}).  For an output-error-rate of $10^{-15}$,
the NCG of the I.5 code is $8.5$ dB, which is $1.47$ dB from capacity.

Finally, in Appendix I.9 of G.975.1, a product-like code with
$(1020,988)$ doubly-extended binary BCH component codes is proposed.
The overall code is described in terms of a $512 \times 1020$ matrix
of bits, in which the bits along both the rows of the matrix as well
as a particular choice of `diagonals' must form valid codewords in the
component code. Since the diagonals are chosen to include 2 bits in
every row, any diagonal codeword has two bits in common with any row
codeword; in contrast, for a product code, any row and column have
exactly one bit in common. Note that the I.9 construction achieves a
product-like construction (their choice of diagonals ensures that each
bit \emph {is} protected by two component codewords) with essentially
half the overall block length of the related product code (even so,
the I.9 code has the longest block length among all G.975.1
proposals). However, the choice of diagonals decreases the size of the
smallest stall patterns, introducing an error floor above $10^{-14}$.
For an output-error-rate of $2\cdot10^{-14}$, the NCG of the I.9 code
is $8.67$ dB, which is $1.3$ dB from capacity.  

\section{LDPC~vs.~Product Codes} \label{sec:vs}

In this section, we present a high-level view of iterative decoders
for LDPC and product codes.  Due to the differences in their
implementations, a precise comparison of their implementation
complexities is difficult.  Nevertheless, since the communication
complexity of message-passing is a significant challenge in LDPC
decoder design, we consider the decoder data-flow, i.e., the rate of
routing/storing messages, as a surrogate for the implementation
complexity.   

\subsection{Decoder-Data-flow Comparison}

We consider a system that transmits \emph{information} at $D$ bits/s,
using a binary error-correcting code of rate $R$---for which
\emph{hard} decisions at $D/R$ bits/s are input to the decoder---and a
decoder that operates at a clock frequency $f_c$ Hz. 

\subsubsection{LDPC Code}

We consider an LDPC decoder that implements sum-product decoding (or
some quantized approximation) with a parallel-flooding schedule. We
assume $q$-bit messages internal to the decoder, an average variable
node degree $d_{av}$, and $N$ decoder iterations; typically, $q$ is 4
or 5 bits, $d_{av}\approx 3$, and $N\sim15-25$.  Initially,
hard-decisions are input to the decoder at a rate of $D/R$ bits/s and
stored in flip-flop registers. At each iteration, variable nodes
compute and broadcast $q$-bit messages over every edge, and similarly
for the check nodes, i.e., $2qd_{av}$ bits are broadcast per iteration
per variable node. Since bits arrive from the channel at $D/R$ bits/s,
the corresponding internal data-flow per iteration is then
$D2qd_{av}/R$, and the total data-flow, including initial loading of
1-bit channel messages, is
\begin{IEEEeqnarray*}{rCl}
F_{\rm LDPC}&=&\frac{D}{R}+\frac{2 N D q d_{av}}{R} \\
&\approx & \frac{2 N D q d_{av}}{R}.
\end{IEEEeqnarray*}

For $N=20$, $q=4$, $d_{av}=3$, $F_{\rm LDPC}\approx 480D/R$, which
corresponds to a data-flow of more than $48$ Tb/s for $100$ Gb/s
systems.

\begin{figure}[!t]
\centering
\includegraphics[scale=0.5]{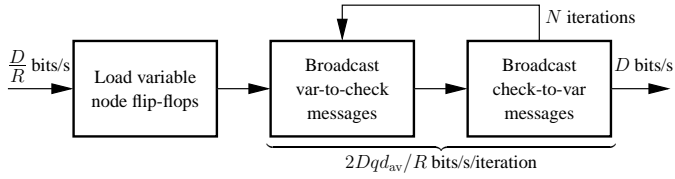}
\caption{Data-flow in an LDPC decoder} \label{fig:dflowldpc}
\end{figure}

\subsubsection{Product Code} \label{sec:proddec}

When the component codes of a product code can be efficiently decoded
via syndromes (e.g., BCH codes), there exists an especially efficient
decoder for the product code. Briefly, by operating exclusively in the
`syndrome domain'---which compresses the received signal---and passing
only $\leq t$ messages per (component) decoding (for
$t$-error-correcting component codes), the implementation complexity
of decoding is significantly reduced. 

The following is a step-by-step description of the decoding algorithm:
\begin{enumerate}
\item	From the received data, compute and store the syndrome for
each row and column codeword. Store a copy of the received data in
memory $R$.	
\item	Decode those non-zero syndromes corresponding to row
codewords\footnote{In practice, the syndrome corresponding to a fixed
row is decoded only if its value has changed since its last
decoding.}. In the event of a successful decoding, set the syndrome to
zero, flip the corresponding $t$ or fewer positions in memory $R$, and
update the $t$ or fewer affected column syndromes by a masking
operation.
\item Repeat Step 2, reversing the roles of rows and columns.
\item If any syndromes are non-zero, and fewer than the maximum number
of iterations have been performed, go to Step 2. Otherwise, output the
contents of memory $R$.
\end{enumerate}

We quantify the complexity of decoding a product code by its decoder
data-flow.  At first glance, it may seem that this approach ignores
the complexity of decoding the (component) $t$-error-correcting BCH
codewords. However, for relatively small $t$, the decoding of a
component codeword can be efficiently decomposed into a series of
look-up table operations, for which the data-flow interpretation is
well-justified.  In this section, we will ignore the data-flow
contribution of the BCH decoding algorithm, but we return to this
point in the Appendix, where it is shown that the corresponding
data-flow is negligible.

We assume that rows are encoded by a $t_1$-error-correcting
$(n_1,k_1=n_1-r_1)$ BCH code, and the columns are encoded by a
$t_2$-error-correcting $(n_2,k_2=n_2-r_2)$ BCH code, for an overall
rate $R=R_1R_2$.  We assume each row/column codeword is decoded (on
average, over the course of decoding the overall product code) $v$
times, where typically $v$ ranges from 3 to 4.  

\begin{figure}[!t]
\centering
\includegraphics[scale=0.5]{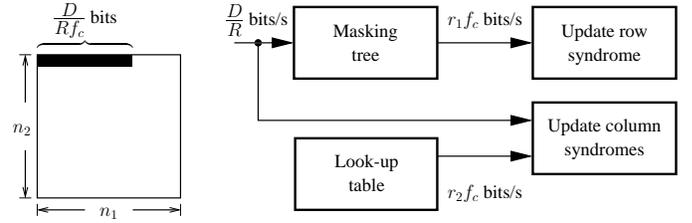}
\caption{Data-flow in the initial syndrome computing} \label{fig:dflowsyn}
\end{figure}

The hard-decisions from the channel---at D/R bits/s---are written to a
data RAM, in addition to being processed by a syndrome
computation/storage device.  Contrary to the LDPC decoder data-flow,
the clock frequency $f_c$ plays a central role, namely in the
data-flow of the initial syndrome calculation. Referring to
Fig.~\ref{fig:dflowsyn}, and assuming that the bits in a product code
are transmitted row-by-row, the input bus-width (i.e., the number of
input bits per decoder clock cycle) is $D/(Rf_c)$ bits. Now, assuming
these bits correspond to a single row of the product code, each
non-zero bit corresponds to some $r_1$-bit mask (i.e., the
corresponding column of the parity-check matrix of the row code), the
modulo-2 sum of these is performed by a masking tree, and the
$r_1$-bit output is masked with the current contents of the
corresponding (syndrome) flip-flop register.  That is, each clock
cycle causes a $r_1$-bit mask to be added to the contents of the
corresponding row in the syndrome bank.  Of course, each received bit
also impacts a distinct column syndrome, however, the \emph{same}
$r_2$-bit mask is applied (when the corresponding received bit is
non-zero) to each of the involved column syndromes; the corresponding
data-flow is then $r_2$ bits per clock cycle.

Once the syndromes are computed from the received data, iterative
decoding commences. To perform a row decoding, an $r_1$-bit syndrome
is read from the syndrome bank. Since there are $n_2$ row codewords,
and each row is decoded on average $v$ times, the corresponding
data-flow from the syndrome bank to the row decoder is $r_1 n_2 v D/(R
n_1 n_2)=r_1 v D/(R n_1)$ bits/s.  For each row decoding, at most
$t_1$ positions are corrected, each of which is specified by $\lceil
\log_2  n_1 \rceil + \lceil \log_2 n_2 \rceil$ bits.  Therefore, the
data-flow from the row decoder to the data RAM is
\[
\frac{t_1 n_2 v D (\lceil \log_2 n_1 \rceil + \lceil \log_2  n_2
\rceil)}{R n_1 n_2}=\frac{t_1 v D (\lceil \log_2 n_1 \rceil + \lceil
\log_2  n_2 \rceil)}{R n_1}
\]
bits/s. Furthermore, for each corrected bit, a $r_2$-bit mask must be
applied to the corresponding column syndrome, which yields a data-flow
from the row decoder to the syndrome bank of $t_1 n_2 r_2 v D/(R n_1
n_2)=t_1 r_2 v D/(R n_1)$ bits/s. A similar analysis can be applied to
column decodings.  In total, the decoder data-flow is
\begin{IEEEeqnarray*}{rCl}
F_{\rm P}&=&\frac{D}{R}+(r_1+r_2)\cdot f_c\\
& & \mbox{} +\frac{D v}{Rn_1}\cdot \left(t_1  \lceil \log_2 n_1 \rceil
+ t_1 \lceil \log_2  n_2 \rceil+r_1+t_1r_2 \right) \\
& & \mbox{} +\frac{D v}{Rn_2}\cdot \left(t_2 \lceil \log_2  n_1 \rceil
+ t_2 \lceil \log_2  n_2 \rceil+r_2+t_2r_1 \right). 
\end{IEEEeqnarray*}

\begin{figure}[!t]
\centering
\includegraphics[scale=0.5]{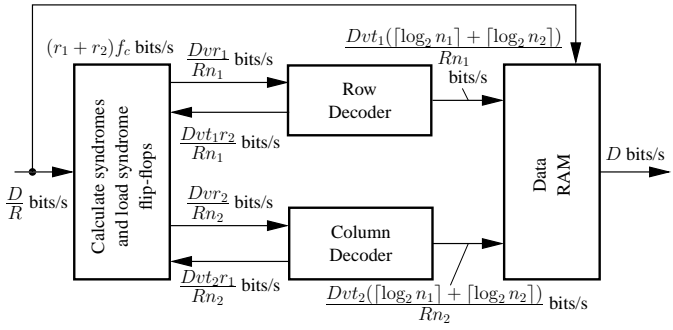}
\caption{Data-flow in a product-code decoder} \label{fig:dflowprod}
\end{figure}

In this work, we will focus on codes for which $n_1=n_2\approx 1000$,
$r_1=r_2=32$, $t_1=t_2=3$, and the decoder is assumed to operate at
$f_c \approx 400$~MHz. For $v=4$, we then have a data-flow of
approximately 293 Gb/s.  Note that this is more than two orders of
magnitude smaller than the corresponding data-flow for LDPC decoding.
Intuitively, the advantage arises from two facts. First, when
$R_1>1/2$ and $R_2>1/2$, syndromes provide a compressed representation
of the received signal.  Second, the algebraic component codes admit
an economical message-passing scheme, in the sense that message
updates are only required for the small fraction of bits that are
corrected by a particular (component code) decoding. 

\section{Staircase Codes} \label{sec:sc}

The staircase code construction combines ideas from recursive
convolutional coding and block coding. Staircase codes are completely
characterized by the relationship between successive matrices of
symbols. Specifically, consider the (infinite) sequence
$B_0,B_1,B_2,\ldots$ of $m$-by-$m$ matrices $B_i$, $i \in
\mathbb{Z}^{+}$. Herein, we restrict our attention to $B_i$ with
elements in $\mathbb{F}_2$, but an analogous construction applies in
the non-binary case.  

Block $B_0$ is initialized to a reference state known to the
encoder-decoder pair, e.g., block $B_0$ could be initialized to the
all-zeros state, i.e., an $m$-by-$m$ array of zero symbols.
Furthermore, we select a conventional FEC code (e.g., Hamming, BCH,
Reed-Solomon, etc.) in systematic form to serve as the
\emph{component} code; this code, which we henceforth refer to as $C$,
is selected to have block length $2m$ symbols, $r$ of which are parity
symbols.   

Encoding proceeds recursively on the $B_i$.  For each $i$, $m(m-r)$
information symbols (from the streaming source) are arranged into the
$m-r$ leftmost columns of $B_i$; we denote this sub-matrix by
$B_{i,L}$.  Then, the entries of the rightmost $r$ columns (this
sub-matrix is denoted by $B_{i,R}$) are specified as follows:

\begin{enumerate}
\item Form the $m \times (2m-r)$ matrix, $A=\left[ B_{i-1}^T  \
B_{i,L} \right]$, where $B_{i-1}^T$ is the matrix-transpose of
$B_{i-1}$.
\item The entries of $B_{i,R}$ are then computed such that each of the
rows of the matrix $\left[ B_{i-1}^T \ B_{i,L} \  B_{i,R} \right]$ is
a valid codeword of $C$. That is, the elements in the $j$th row of
$B_{i,R}$ are exactly the $r$ parity symbols that result from encoding
the $2m-r$ `information' symbols in the $j$th row of $A$.
\end{enumerate}

\begin{figure}[!t] 
\centering
\includegraphics[scale=0.5]{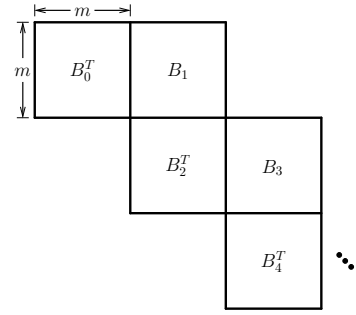}
\caption{The `staircase' visualization of staircase codes.} 
\label{fig:sc} 
\end{figure}

Generally, the relationship between successive blocks in a staircase
code satisfies the following relation: for any $i\geq 1$, each of the
rows of the matrix $\left[ B_{i-1}^T B_i \right]$ is a valid codeword
in $C$. An equivalent description---from which the term `staircase
codes' originates---is suggested by Fig.~\ref{fig:sc}, in which (the
concatenation of the symbols in) every row (and every column) in the
`staircase' is a valid codeword of $C$; this representation suggests
their connection to product codes.  However, staircase codes are
naturally unterminated (i.e., their block length is indeterminate),
and thus admit a range of decoding strategies with varying latencies.
Most importantly, we will see that they outperform product codes.

The rate of a staircase code is
\[
R_{\rm s}=1-\frac{r}{m},
\]
since encoding produces $r$ parity symbols for each set of $m-r$ `new'
information symbols.  However, note that the related product code has
rate
\begin{IEEEeqnarray*}{rCl}
R_{\rm p}&=& \left( \frac{2m-r}{2m} \right )^2 \\
&=& 1-\frac{r}{m}+\frac{r^2}{4m^2},
\end{IEEEeqnarray*}
which is greater than the rate of the staircase code. However, for
sufficiently high rates, the difference is small, and staircase codes
outperform product codes of the same rate. 

From the context of transmitter latency---which includes encoding
latency and frame-mapping latency---staircase codes have the advantage
(relative to product codes) that the \emph{effective} rate (i.e., the
ratio of `new' information symbols, $m-r$, to the total number of
`new' symbols, $m$) of a component codeword is exactly the rate of the
\emph{overall} code.  Therefore, the encoder produces parity at a
`regular' rate, which enables the design of a frame-mapper that
minimizes the transmitter latency.

We note that staircase codes can be interpreted as generalized LDPC
codes with a \emph{systematic} encoder and an indeterminate
block-length, which admits decoding algorithms with a range of
latencies.

Using arguments analogous to those used for product codes, a
$t$-error-correcting component code $C$ with minimum distance
$d_{\rm{min}}$ has a Hamming distance between any two staircase
codewords that is at least $d_{\rm{min}}^2$. 

\subsection{Decoding Algorithm}

Staircase codes are naturally unterminated (i.e., their block length
is indeterminate), and thus admit a range of decoding strategies with
varying latencies. That is, decoding can be accomplished in a
sliding-window fashion, in which the decoder operates on the received
bits corresponding to $L$ consecutively received blocks $B_i, B_{i+1},
\ldots, B_{i+L-1}$. For a fixed $i$, the decoder iteratively decodes
as follows: First, those component codewords that `terminate' in block
$B_{i+L-1}$ (i.e., whose parity bits are in $B_{i+L-1}$) are decoded;
since every symbol is involved in two component codewords, the
corresponding syndrome updates are performed, as in
Section~\ref{sec:proddec}. Next, those codewords that terminate in
block $B_{i+L-2}$ are decoded. This process continues until those
codewords that terminate in block $B_i$ are decoded.  Now, since
decoding those codewords terminating in some block $B_j$ affects those
codewords that terminate in block $B_{j+1}$, it is beneficial to
return to $B_{i+L-1}$ and to repeat the process. This iterative
process continues until some maximum number of iterations is
performed, at which time the decoder outputs its estimate for the
contents of $B_i$, accepts in a new block $B_{i+L}$, and the entire
process repeats (i.e., the decoding window slides one block to the
`right').

\subsection{Multi-edge-type Interpretation}

Staircase codes have a simple graphical representation, which provides
a multi-edge-type~\cite{RU2008a} interpretation of their construction.
The term `multi-edge-type' was originally applied to describe a
refined class of irregular LDPC codes, in which variable nodes (and
check nodes) are classified by their degrees with respect to a set of
edge types. Intuitively, the introduction of multiple edge types
allows degree-one variable nodes, punctured variable nodes, and other
beneficial features that are not admitted by the conventional
irregular ensemble. In turn, better performance for finite
blocklengths and fixed decoding complexities is possible.

\begin{figure}[!t] 
\centering
\includegraphics[width=\columnwidth]{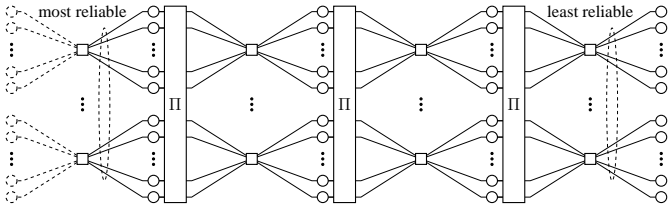}
\caption{A multi-edge-type graphical representation of staircase codes. $\Pi$ is a standard block interleaver, i.e., it represents
the transpose operation on an $m$-by-$m$ matrix.} 
\label{fig:fg} 
\end{figure}

In Fig.~\ref{fig:fg}, we present the factor graph representation of a
decoder that operates on a window of $L=4$ blocks; the graph for
general $L$ follows in an obvious way.  Dotted variable nodes indicate
symbols whose value was decoded in the previous stage of decoding.
The key observation is that when these symbols are correctly
decoded---which is essentially always the case, since the output BER
is required to be less than $10^{-15}$---the component codewords in
which they are involved are effectively shortened by $m$ symbols.
Therefore, the most reliable messages are passed over those edges
connecting variable nodes to the shortened (component) codewords, as
indicated in Fig.~\ref{fig:fg}. On the other hand, the rightmost
collection of variable nodes are (with respect to the current decoding
window) only involved in a single component codeword, and thus the
edges to which they are connected carry the least reliable messages.
Due to the nature of iterative decoding, the intermediate edges carry
messages whose reliability lies between these two extremes. 

\subsection{A G.709-compatible Staircase Code} \label{sec:scG709}

The ITU-T Recommendation G.709 defines the framing structure and
error-correcting coding rate for OTNs. For our purposes, it suffices
to know that an optical frame consists of $130560$ bits, $122368$ of
which are information bits, and the remaining $8192$ are parity bits,
which corresponds to error-correcting codes of rate $R=239/255$.
Since $(510-32)/510=239/255$, we will consider a component code with
$m=510$ and $r=32$. Specifically, the binary $(n=1023,k=993,t=3)$ BCH
code with generator polynomial
$(x^{10}+x^{3}+1)(x^{10}+x^3+x^2+x+1)(x^{10}+x^8+x^3+x^2+1)$ is
adapted to provide an additional 2-bit error-detecting mechanism,
resulting in the generator polynomial\footnote{This is the code
applied to the rows (but not the slopes) of the I.9 code in G.975.1.} 
\begin{IEEEeqnarray*}{rCl}
g(x)&=&(x^{10}+x^{3}+1)(x^{10}+x^3+x^2+x+1)  \\
& & \mbox{} \cdot (x^{10}+x^8+x^3+x^2+1)(x^2+1).
\end{IEEEeqnarray*}
In order to provide a simple mapping to the G.709 frame, we first note
that $2 \cdot 130560=510 \cdot 512$. This leads us to define a slight
generalization of staircase codes, in which the blocks $B_i$ consist
of $512$ rows of $510$ bits. The encoding rule is modified as follows:

\begin{enumerate}
\item Form the $512 \times (512+510) $ matrix, $A=\left[
\hat{B}_{i-1}^T  \ B_{i,L} \right]$, where $\hat{B}_{i-1}^T$ is
obtained by appending two all-zero rows to the top of the
matrix-transpose of $B_{i-1}$.
\item The entries of $B_{i,R}$ are then computed such that each of the
rows of the matrix $\left[ B_{i-1}^T \ B_{i,L} \  B_{i,R} \right]$ is
a valid codeword of $C$. That is, the elements in the $j$th row of
$B_{i,R}$ are exactly the $32$ parity symbols that result from
encoding the $990$ `information' symbols in the $j$th row of $A$.
\end{enumerate}

Here, $C$ is the code obtained by shortening the code generated by
$g(x)$ by one bit, since our overall codeword length is
$510+512=1022$.

\section{Error Floor Analysis} \label{sec:efloor}

\begin{figure}[!t] 
\centering
\includegraphics[scale=0.5]{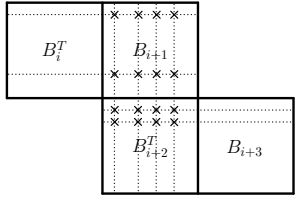}
\caption{A stall pattern for a staircase code with a triple-error correcting component code. Since every involved component codeword has 4 errors, decoding stalls.} 
\label{fig:st} 
\end{figure}

For iteratively decoded codes, an error floor (in the output
bit-error-rate) can often be attributed to error patterns that
`confuse' the decoder, even though such error patterns could easily be
corrected by a maximum-likelihood decoder.  In the context of LDPC
codes, these error patterns are often referred to as trapping
sets~\cite{rich2003}. In the case of product-like codes with an
iterative hard-decision decoding algorithm, we will refer to them as
\emph{stall patterns}, due to the fact that the decoder gets locked in
a state in which no updates are performed, i.e., the decoder stalls,
as in Fig.~\ref{fig:st}.  
\begin{defn}
A stall pattern is a set $s$ of codeword positions, for which every
row and column involving positions in $s$ has at least $t+1$ positions
in $s$.
\end{defn}
We note that this definition includes stall patterns that \emph{are}
correctable, since an \emph{incorrect} decoding may fortuitously cause
one or more bits in $s$ to be corrected, which could then lead to all
bits in $s$ eventually being corrected.  In this section, we obtain an
estimate for the error floor by over-bounding the probabilities of
these events, and pessimistically assuming that every stall pattern is
uncorrectable (i.e., if any stall pattern appears during the course of
decoding, it will appear in the final output).  The methods presented
for the error floor analysis apply to a general staircase code, but
for simplicity of the presentation, we will focus on a staircase code
with $m=510$ and doubly-extended triple-error-correcting component
codes.

\subsection{A Union Bound Technique}

Due to the streaming nature of staircase codes, it is necessary to
account for stall patterns that span (possibly multiple) consecutive
blocks.  In order to determine the bit-error-rate due to stall
patterns, we consider a fixed block $B_i$, and the set of stall
patterns that include positions in $B_i$.  Specifically, we `assign'
to $B_i$ those stall patterns that include symbols in $B_i$ (and
possibly additional positions in $B_{i+1}$) but no symbols in
$B_{i-1}$.  Let $\mathcal{S}_i$ represent the set of stall patterns
assigned to $B_i$. By the union bound, we then have
\[
{\rm BER}_{\rm floor} \leq \sum_{s \in \mathcal{S}_i} \mbox{Pr}[\mbox{bits in } s \mbox{ in error}]\cdot \frac{|s|}{510^2}.
\]
Therefore, bounding the error floor amounts to enumerating the set
$\mathcal{S}_i$, and evaluating the probabilities of its elements
being in error.

\subsection{Bounding the Contribution Due to Minimal Stalls}

\begin{defn}
A minimal stall pattern has the property that there are only $t+1$
rows with positions in $s$, and only $t+1$ columns with positions in
$s$.
\end{defn}

The minimal stall patterns of a staircase code can be counted in a
straightforward manner; the multiplicity of minimal stall patterns
that are assigned to $B_i$ is 
\[
M_{\rm min}=\binom{510}{4}\cdot \sum_{m=1}^4 \binom{510}{m} \cdot \binom{510}{4-m},
\]
and we refer to the set of minimal stall patterns by $\mathcal{S}_{\rm
min}$.  The probability that the positions in some minimal stall
pattern $s$ are \emph{received} in error is $p^{16}$.

Next, we consider the case in which not all positions in some minimal
stall pattern $s$ are received in error, but that due to incorrect
decoding(s), all positions in $s$ are---at some point during
decoding---simultaneously in error.  For some fixed $s$ and $l$,
$1\leq l \leq16$, there are  $\binom{16}{l}$ ways in which $16-l$
positions in $s$ can be received in error. For the moment, let's
assume that erroneous bit flips occur independently with some
probability $\zeta$, and that $\zeta$ does not depend on $l$.   Then
we can \emph{overbound} the probability that a \emph{particular}
minimal stall $s$ occurs by
\[
\sum_{l=0}^{16} \binom{16}{l} p^{16-l}
\zeta^l =(p+\zeta)^{16}.
\]

In order to provide evidence in favor of these assumptions,
Table~\ref{tab:stall} presents empirical estimates, for $l=0$, $l=1$
and $l=2$, of the probability that a minimal stall pattern $s$ occurs
during iterative decoding, given that $16-l$ positions in $s$ are
(intentionally) received in error.  Note that even if a minimal stall
is received, there exists a non-zero probability that it will be
corrected as a result of erroneous decodings; we will ignore this
effect in our estimation, i.e., we make the worst-case assumption that
any minimal stall persists.  Furthermore, from the results for $l=1$
and $l=2$, it appears that our stated assumptions regarding $\zeta$
hold true, and $\zeta\approx 5.8\times 10^{-4}$. For $l>2$, we did not
have access to sufficient computational resources for estimating the
corresponding probabilities. Nevertheless, based on the evidence
presented in Table~\ref{tab:stall}, the error floor contribution due
to minimal stall patterns is estimated as
\[
\frac{16}{510^2} \cdot M_{\rm min} \cdot (p+\zeta)^{16},
\]
where $\zeta=5.8\times 10^{-4}$ when $p=4.8\times 10^{-3}$.

\begin{table}[ht!]
\centering
\caption{Estimated probability of a minimal stall $s$, given that $16-l$
positions are received in error}
\label{tab:stall}
\begin{tabular}{lc} 
$l$ & Estimated probability \\ \hline
0 & $149/150$ \\ 
1 & $1/1725$ \\ 
2 & $(1/1772)^2$ \\\hline
\end{tabular}
\end{table}

\subsection{Bounding the Contribution Due to Non-minimal Stalls}

\begin{figure}[!t] 
\centering
\includegraphics[scale=0.5]{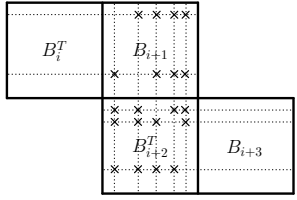}
\caption{A non-minimal stall pattern for a staircase code with a
triple-error correcting component code.} 
\label{fig:KLstall} 
\end{figure}

We now wish to account for the error floor contribution of non-minimal
stalls, e.g., the stall pattern illustrated in Fig.~\ref{fig:KLstall}.
In the general case, a stall pattern $s$ includes codeword positions
in $K$ rows and $L$ columns, $K\geq 4$, $L\geq 4$; we refer to these
as $(K,L)$-stalls.  Furthermore, each $(K,L)$-stall includes $l$
positions, $4\cdot\max(K,L) \leq l \leq K \cdot L$, where the lower
bound follows from the fact that every row and column (in the stall)
includes at least 4 positions.  Note that there are 
\[
A_{K,L}=\binom{510}{L} \cdot \sum_{m=1}^K \binom{510}{m} \cdot
\binom{510}{K-m}
\]
ways to select the involved rows and columns. 

For a fixed $(K,L)\neq(4,4)$ and a fixed choice of rows and columns,
we now proceed to overbound the contributions of candidate stall
patterns.  Without loss of generality, we assume that $K \geq L$, and
note that there are
${\binom{L}{4}}^{K}$
ways of choosing $l=4K$ elements (in the $L\cdot K$ `grid' induced by
the choice of rows and columns) such that each column includes exactly
four elements, and that every stall pattern `contains' at least one of
these. Now, since a stall pattern includes $l$ elements, $4\cdot K
\leq l \leq K \cdot L$, the number of stall patterns with $l$ elements
is overbounded as
\[
{\binom{L}{4}}^{K}\cdot\binom{K\cdot L-4\cdot K}{l-4\cdot K}.
\]

For a general $(K,L)\neq (4,4)$, it follows that the number of stall
patterns with $l$ elements, $4\cdot \max (K,L) \leq l \leq K\cdot L$,
is overbounded as
\[
{\binom{\min (K,L)}{4}}^{\max (K,L)}\cdot\binom{K\cdot L-4\cdot \max (K,L)}{l-4\cdot \max (K,L)}.
\]
Finally, over the choice of the $K$ rows and $L$ columns, there are
\[M_{K,L}^l= A_{K,L} \cdot{\binom{\min (K,L)}{4}}^{\max
(K,L)}\cdot\binom{KL-4\cdot \max (K,L)}{l-4\cdot \max (K,L)}\]
$(K,L)$-stalls with $l$ elements.

For a fixed $K$ and $L$, the contribution to the error floor can be
estimated as
\[
\sum_{l=4\cdot\max(K,L)}^{K \cdot L} \frac{l}{510^2} \cdot M_{K,L}^l \cdot (p+\zeta)^{l},
\]
and in Table~\ref{tab:KL}, we provide values for various $K$ and $L$,
when $\zeta=5.8\times 10^{-4}$ and $p=4.8\times 10^{-3}$.  

\begin{table}[!b]
\centering
\caption{Contribution to Error Floor Estimate of $(K,L)$-stall patterns}
\label{tab:KL}
\begin{tabular}{ccc}
$K$ & $L$ & Contribution \\ \hline
4 & 4 & $3.55\times 10^{-21}$ \\
4 & 5 & $7.81\times 10^{-28}$ \\
5 & 5 & $2.54\times 10^{-22}$ \\
5 & 6 & $2.21\times 10^{-28}$ \\
6 & 6 & $1.40\times 10^{-23}$ \\
6 & 7 & $1.49\times 10^{-29}$ \\
7 & 7 & $8.53\times 10^{-25}$ \\
7 & 8 & $1.83\times 10^{-32}$ \\\hline
\end{tabular}
\end{table}

Note that the dominant contribution to the error floor is due to
minimal stall patterns (i.e., $K=L=4$), and that the overall estimate
for the error floor of the code is $3.8\times 10^{-21}$. Finally, we
note that by a similar (but more cumbersome) analysis, the error floor
of the G.709-compliant staircase code is estimated to occur at
$4.0\times 10^{-21}$.

\section{Simulation Results} \label{sec:sim}

\begin{figure}[!t] 
\centering
\includegraphics[width=\columnwidth]{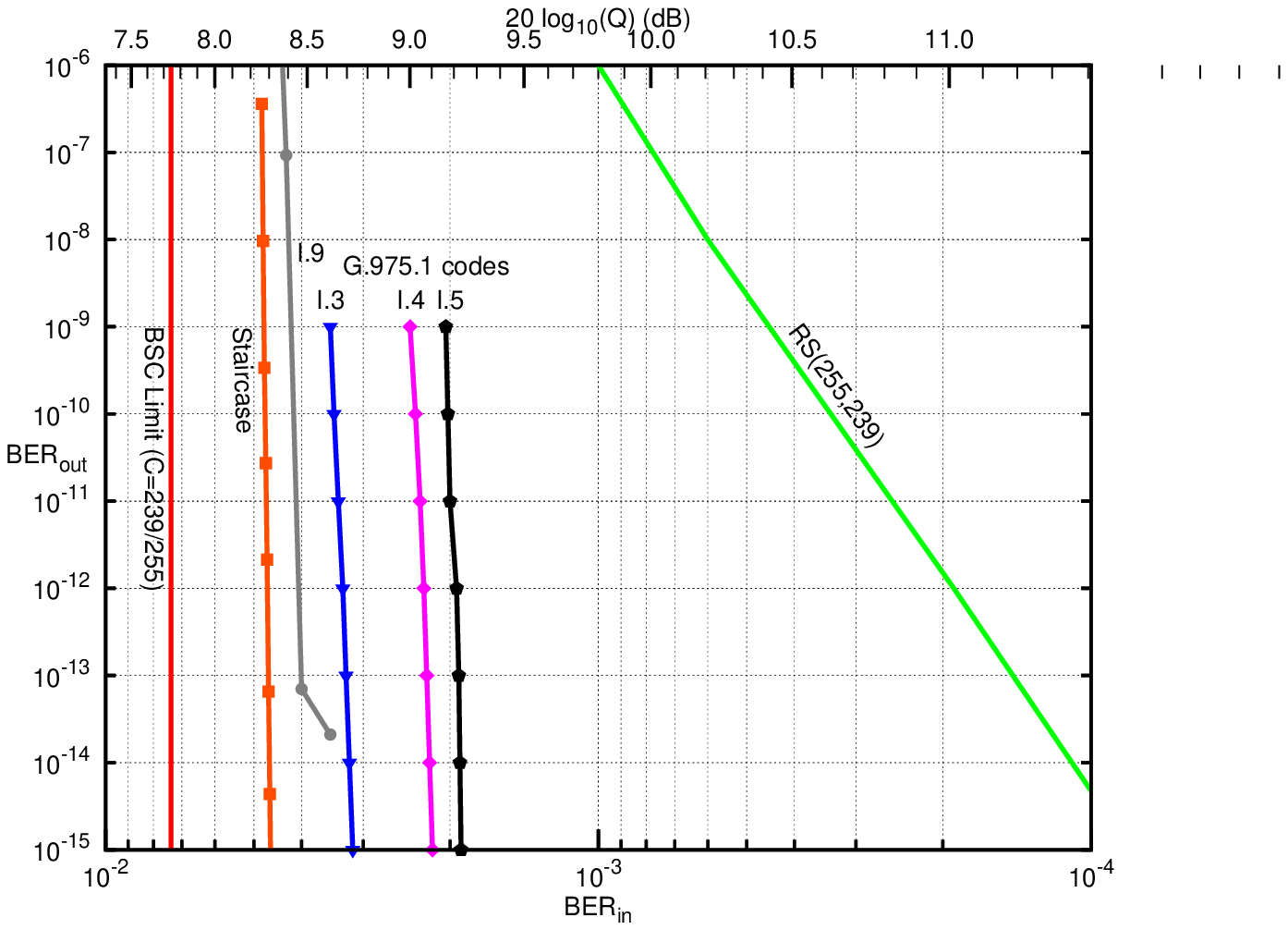}
\caption{Performance of a $R=239/255$ staircase code on a
binary symmetric channel with crossover probability
$\mathrm{BER}_{\mathrm{in}}$, compared
with various G.975.1 codes.  The upper scale plots the equivalent
binary-input Gaussian channel $Q$ (in dB),
where $\mathrm{BER}_{\mathrm{in}}=(1/2)\mathrm{erfc}(Q/\sqrt{2})$.}
\label{fig:ber} 
\end{figure}

In Fig.~\ref{fig:ber}, simulation results---generated in hardware on
an FPGA implementation---are provided for the G.709-compatible
staircase code, for $L=7$.  We also present the bit-error-rate curves
for the G.975 RS code, as well as the G.975.1 codes described in
Section~\ref{sec:exist}.  For an output error rate $10^{-15}$, the
staircase code provides approximately $9.41$ dB net coding gain, which
is within $0.56$ dB of the Shannon limit, and an improvement of $0.42$
dB relative to the best G.975.1 code.

\section{Conclusions}

We proposed staircase codes, a class of product-like FEC codes that
provide reliable communication for streaming sources.  Their
construction admits low-latency encoding and variable-latency
decoding, and a decoding algorithm with an efficient hardware
implementation.  For $R=239/255$, a G.709-compatible staircase code
was presented, and performance within $0.56$~dB of the Shannon Limit
at $10^{-15}$ was provided via an FPGA-based simulation.

\appendix
This section briefly describes known techniques for efficiently
decoding triple-error-correcting binary BCH codes, and discusses the
data-flow associated with a lookup-table-based decoder architecture. 

For a syndrome $\mathbf{S}=(S_1,S_3,S_5)$, $S_i \in \mathbb{F}_{2^m}$,
we first compute $D_3  =  S_1^3 + S_3$ and $D_5 = S_1^5 + S_5$.  A
triple-error correcting decoder distinguishes the cases
\[
\begin{array}{ll}
v = 0: & S_1 = S_2 = S_3 = 0 \\
v = 1: & S_1 \neq 0, D_3 = D_5 = 0 \\
v = 2: & S_1 \neq 0, D_3 \neq 0, S_1 D_5 = S_3 D_3 \\
v = 3: & D_3 \neq 0, v \neq 2,
\end{array}
\]
where $v$ is the number of positions to invert in order to obtain a
valid codeword.

In order to determine the  corresponding positions, a reciprocal
error-locator polynomial $\tilde{\sigma}(x)$ is defined, the roots of
which identify the positions. From~\cite{Ree1999a}, we have:
\[
\begin{array}{ll}
v = 1: &  \tilde{\sigma}(x) = x + S_1 \\
v = 2: &  \tilde{\sigma}(x) = x^2 + S_1 x + D_3/S_1 \\
v = 3: &  \tilde{\sigma}(x) = x^3 + S_1 x^2 + bx +  S_1 b + D_3
\end{array}
\]
where
\[
b = (S_1^2 S_3 + S_5)/D_3.
\]
When $t=2$, note that all of the coefficients of $\tilde{\sigma}(x)$
are nonzero.

It remains to determine the roots of $\tilde{\sigma}(x)$.  For $v=1$,
it is trivial to determine the error location.  For $v=2$ or $v=3$,
lookup-based methods for solving the corresponding quadratic and cubic
equations are described in \cite{chi1969a,ber1967a}. In the remainder
of this section, we briefly describe these methods, and discuss their
data-flow.

For a quadratic equation $f_X(x) = x^2 + ax + b$ with $a \neq 0$,
substitute $x = ay$ to obtain
\[
f_Y(y) = a^2( y^2 + y + b/a^2).
\]
If $f_Y(r) = 0$ then $f_X(ar) = 0$.  Thus the problem of finding roots
of $f_X(x)$ reduces to the problem of finding roots of the suppressed
quadratic $f_Y(y)$, which can be solved by lookup using a table with
$2^m$ entries, each of which is a pair of elements in
$\mathbb{F}_{2^m}$.  Therefore, when $v=2$, decoding requires $2m$
bits to be read from a lookup-table memory.

Similarly, for a cubic equation $f_X(x) = x^3 + ax^2 + bx + c$,
substitute $x = y+a$ to obtain
\[
f_Y(y) = y^3 + (a^2 + b)y + ab+c.
\]

Note that $yf_Y(y)$ is a linearized polynomial with respect to $\mathbb{F}_2$
and hence the set of zeros of $yf_Y(y)$ is a vector space over $\mathbb{F}_2$.
In particular, the roots of $y f_Y(y)$, if distinct,  are of the form
$\{ 0, r_1, r_2, r_1 + r_2 \}$.  Thus only $r_1$ and $r_2$ need to be
stored in the lookup table.  

Two cases arise, depending on the value of $a^2 + b = D_5/D_3$.  If
$D_5 = 0$, so that $a^2 + b = 0$, then $f_Y(y) = y^3 + ab + c$, and
the roots can be found by finding the cube roots of $ab + c= D_3$,
which requires lookup using a table with $2^m$ entries, each of which
is a pair of elements in $\mathbb{F}_{2^m}$.  If $D_5 \neq 0$, so that
$a^2 + b \neq 0$, substitute $y = (a^2 + b)^{1/2}z$ to obtain
\[
f_Z(z) = (a^2 + b)^{3/2} (z^3 + z + (ab+c)/(a^2 + b)^{3/2}),
\]
where
\[
\frac{ab+c}{(a^2+b)^{3/2}} = \left( \frac{D_3^5}{D_5^3}
\right)^{1/2}.
\]
\newpage
The roots of the suppressed cubic $f_Z(z)$ can be
found by lookup using a table with $2^m$ entries, each of which is a
pair of elements in $\mathbb{F}_{2^m}$.  Therefore, in either case,
decoding requires $2m$ bits to be read from a lookup-table memory.

Finally, for $n=n_1=n_2$, the data-flow contribution of the
lookup-table-based decoding architecture is $\frac{4mvD}{nR}$.  For
$n=1000$, $m=10$, $v=4$, $R=239/255$ and $D=100$ Gb/s, the
corresponding data-flow is $17.1$ Gb/s, which is small relative to the
data-flow that arises due to those effects considered in
Section~\ref{sec:proddec}.

\end{document}